\documentclass[12pt]{article}
\usepackage{a4}
\usepackage{epsf}
\global\arraycolsep=2pt %reduces the separation in eqnarrays
\def\HQQET{HQ$\overline{\rm Q}$ET }
\makeatletter
\def\fmslash{\@ifnextchar[{\fmsl@sh}{\fmsl@sh[0mu]}}
\def\fmsl@sh[#1]#2{%
  \mathchoice
    {\@fmsl@sh\displaystyle{#1}{#2}}%
    {\@fmsl@sh\textstyle{#1}{#2}}%
    {\@fmsl@sh\scriptstyle{#1}{#2}}%
    {\@fmsl@sh\scriptscriptstyle{#1}{#2}}}
\def\@fmsl@sh#1#2#3{\m@th\ooalign{$\hfil#1\mkern#2/\hfil$\crcr$#1#3$}}
\makeatother
\begin{document}
%---------------- CERN Titlepage <---------------------------
\thispagestyle{empty}
\begin{titlepage}

\begin{flushright}
TTP 97--02 \\
hep-ph/9701324 \\
\today
\end{flushright}

\vspace{0.3cm}
\boldmath
\begin{center}
\Large\bf QCD based Calculation of the semi-inclusive Decay 
          $\eta_{_Q} \to \gamma$ + light Hadrons 
\end{center}
\unboldmath
\vspace{0.8cm}

\begin{center}
{\large Thomas Mannel} and {\large Stefan Wolf} \\
{\sl Institut f\"{u}r Theoretische Teilchenphysik,
     D -- 76128 Karlsruhe, Germany.} 
\end{center}

\vspace{\fill}

\begin{abstract}
\noindent
A QCD based calculation of the photon spectrum in semi-inclusive
$\eta_Q$ decays is performed. The method applied is an effective 
theory based on a $1/m_Q$ expansion of QCD. 
\end{abstract}
\end{titlepage}

%---------------- END CERN Titlepage <---------------------------
%
\section{Introduction}
The expansion of QCD in inverse powers of the mass has become 
an indispensable tool for the analysis of heavy hadron decays. This 
method, formulated as an effective field theory known as Heavy 
Quark Effective Theory (HQET) \cite{SV87}, has put the physics
of heavy quarks on a QCD based and thus model independent framework.

While applications of these methods to systems with a single heavy 
quark has been studied extensively (see \cite{reviews} for reviews),
there is a lot of motivation to extend this ideas to systems with 
two  or more heavy quarks, because heavy quarkonia 
(the $\psi$'s and the $\Upsilon$'s) have been known since some 
time and the observation of flavored ``doubly heavy'' systems
(such as the $B_c$ family of states) is very likely in the near 
future. 

The starting point of the construction of an effective theory 
for these systems is again the expansion
of the relevant Greens functions in $1/m_Q$, where 
$m_Q$ is the mass of the heavy quark. However, it turns out 
\cite{KMO} that one can not use the static limit for
two heavy quarks, in particular if their velocities differ 
only by an amount of order 
$1/m_Q$, i.e. $vv' - 1 \sim \Lambda_{QCD} / m_Q$ as it is the 
case in a particle containing two heavy quarks. 
The static limit breaks down and one is forced to include at 
least the kinetic energy into the leading order dynamics.

However, unlike the case of one heavy quark an effective theory 
for heavy quarkonia in terms of the quark and gluon degrees of 
freedom is more complicated. This is mainly due to the fact 
that in a heavy quarkonium not only the scale of the light degrees 
of freedom (as e.g. $\bar\Lambda$ in HQET) appears as a scale 
small compared to the heavy quark mass, but also scales which do 
depend on the heavy quark mass. This point becomes clear if one 
studies a system like the positronium first. 

In the QED case one may safely assume that the coupling $\alpha$ does 
not run when one studies a QED bound state consisting of two heavy 
fermions, which we shall assume to be of the same mass $m_Q$. The 
coupling $\alpha$ is a small parameter and hence the inverse Bohr radius 
$m_Q \alpha$ of such a system is small compared to the heavy mass $m_Q$. 
Furthermore, in the case of QED it makes sense to talk about the 
binding energy $E_B$ of this system, which is again smaller by a 
factor $\alpha$, $E_B \sim m_Q \alpha^2$. 

Thus in a coulombic system there are two scales, the inverse Bohr radius
and the binding energy which are both small compared to the heavy mass:
\begin{equation} \label{coul}
m_Q \alpha^2 \ll m_Q \alpha \ll m_Q .
\end{equation}
In addition, both scales depend
on the heavy mass, in this simple example in a trivial way. The expansion 
parameter in this case is the relative velocity $v_{rel}$ between the two
heavy constituents, and a consistent scheme for the velocity counting has
been set up. The velocity in a coulombic system turns out to be 
$v_{rel}= \alpha$ and hence is indeed a small parameter to expand in.
In particular, the binding energy is of the order $v_{rel}^2$. 

Switching now to full QCD the coupling starts to run, introducing another
scale $\Lambda_{QCD}$. The problem in applying the coulombic hierarchy
of scales to realistic quarkonia is related to the question at which point
in the hierarchy (\ref{coul}) $\Lambda_{QCD}$ enters. This is obviously a 
question how large $m_Q$ is, and for ``superheavy'' quarkonia one can 
have the scenario $\Lambda_{QCD} \ll m_Q \alpha^2$ such that the binding 
is coulombic and -- from the point of view of QCD -- is of perturbative 
nature, since $\alpha_s (E_B) \ll 1$. This case is governed by 
Non-Relativistic QCD (NRQCD) as formulated by Bodwin, Braaten and Lepage
\cite{BBL}, where the velocity counting of the QED analogue is used. In 
fact, this type of counting is also correct as long as the binding is 
due to instantaneous gluons such that the emission of a dynamical, transverse
gluon is suppressed by some power of the velocity.

Realistic quarkonia, such as charmonium of bottomonium, do not seem to fulfill
the requirement of being coulombic. This becomes obvious from the spectra 
as well as from the leptonic widths, which are related to the wave functions 
at the origin in the simplest picture. Here it seems that $\Lambda_{QCD}$ 
is at least larger than the binding energy or maybe even larger than the 
inverse Bohr radius. In such a case it is not obvious what the parameter is 
in which an expansion should be formulated. The $1/m_Q$ expansion yields in 
the numerators combinations of covariant derivatives, some of which may be 
identified with the relative velocity. In a ``minimalistic'' approach (called 
\HQQET in \cite{MS}), all derivatives are kept at the price of a 
proliferation of unknown matrix elements. In other words, in \HQQET simply 
the powers of the operators are counted and one can translate
NRQCD into \HQQET and vice versa, finding that some matrix elements in 
\HQQET at leading order are down by factors of $v_{rel}$ in the counting scheme
of NRQCD. 

In the present paper we shall study a simple process which could shed some 
light on the question of how to correctly count the ``powers'' of a matrix 
element in the $1/m_Q$ expansion for realistic heavy quarkonia. We shall 
discuss the decay
$\eta_Q \to \gamma$ + {\it light hadrons}, in which the hard 
subprocess is the annihilation of two heavy quarks 
into a photon and a gluon, which implies that the two heavy 
quarks have to be in a color octet state at the matching scale $m_Q$. 
The matching to the effective theory thus yields 
(at the matching scale $\mu = m_Q$) four-quark operators 
in which the quark-antiquark pair is in a color octet state, 
and hence in naive factorization these would vanish. Also in NRQCD
these matrix elements would require at least an additional dynamical
gluon, the emission of which is suppressed by factors of $v_{rel}$; 
hence this process is an interesting laboratory to test the size of 
these octet contributions which seem to play some role in quarkonia 
production. 
  
In the next section we give a brief review of the effective theory 
approach used here. In section 3 we set up the necessary machinery 
for these decays in which the two heavy quarks annihilate and review 
the short distance contributions. In section 4 we discuss the hadronic 
matrix elements and the restrictions imposed by heavy quarkonia spin 
symmetry. In section 5 we apply this method to calculate the total  
rate and the photon spectrum in $\eta_Q \to \gamma$ + light hadrons. 
Finally we discuss our results and conclude.

\section{Structure of the Effective Theory for Quarkonia}
As we shall exploit the fact that the mass of the heavy quark is large 
compared to both $\Lambda_{QCD}$ and all other scales such as the analoga 
of the inverse Bohr radius  and the binding energy, we are starting from a 
$1/m_Q$ expansion of the QCD Lagrangian and the corresponding 
expansion of the fields up to order $1/m_Q^2$. It is well known that there 
are different ways to perform such an expansion (as e.g. by integrating out 
the small component fields from the QCD functional \cite{MRR} or by a 
Foldy--Wouthuysen Transformation \cite{FW}); although the $1/m_Q$ expansions
of their Lagrangians and of their fields look different, the result for the 
Greens functions will be the same. Hence one may pick the most convenient 
representation for the application under discussion and we pick \cite{FW} \\
\parbox{14cm}{
\begin{eqnarray*}
{\cal L} &=& \bar h_v^{(+)} (iv \cdot D) h_v^{(+)} + K^{(+)}_1 + M^{(+)}_1
           + K^{(+)}_2 + M^{(+)}_2 + {\cal L}_{glue} + \cdots \,,
\\
Q_v^{(+)}(x)
&=& e^{-im_Q v\cdot x}\left[ 1 +\frac{1}{2m_Q}(i\fmslash{D}_{\perp})
+ \frac{1}{4m_Q^2}\left( v\cdot D \fmslash{D}_{\perp}
- \frac{1}{2} \fmslash{D}_{\perp}^2 \right) + \cdots \right] h_v^{(+)}(x) \,,
\end{eqnarray*}
}\hfill\hspace{-2cm}\parbox{1cm}{\begin{equation}\label{FWfield}\end{equation}}
\\
where $v$ is the velocity of the quarkonia, and the transverse components of
the derivative is given by $D^\perp_\mu = (g_{\mu\nu}-v_\mu v_\nu)D^\nu$. The
$K_i$ and $M_i$ are operators of higher dimension
\begin{eqnarray}
K^{(+)}_1 &=& \bar{h}_v^{(+)} \frac{(iD_\perp)^2}{2m_Q} h_v^{(+)} \,, \quad 
M^{(+)}_1 = \frac{1}{2m_Q}
 \bar{h}_v^{(+)} (-i\sigma_{\mu \nu})(iD_\perp^\mu)(iD_\perp^\nu) h_v^{(+)} \,,
\nonumber \\
K^{(+)}_2 &=& \frac{1}{8m_Q^2}
 \bar{h}_v^{(+)} [(iD_\perp^\mu), [(-ivD),(iD^\perp_\mu)]] h_v^{(+)} \,,
\nonumber \\
M^{(+)}_2 &=& \frac{1}{8m_Q^2}
  \bar{h}_v^{(+)} (-i\sigma_{\mu \nu})
                   \{ (iD_\perp^\mu), [(-ivD),(iD_\perp^\nu)] \} h_v^{(+)} \,,
\\
{\cal L}_{glue} &=& 
\left(\frac{1}{2m}\right)^2 \frac{\alpha_s}{30 \pi m^2}
\mbox{ Tr}\{ [ iD_\mu , G^{\mu \nu} ][ iD^\lambda , G_{\lambda \nu} ] \}
\nonumber \\
&& + \left(\frac{1}{2m}\right)^2 \frac{i \alpha_s g_s}{360 \pi m^2}
\mbox{ Tr}\{  G^{\mu \nu} [ G_{\nu \rho} ,
G^{\rho}_{\hskip 0.5em \mu} ] \} \,. \nonumber
\end{eqnarray}
Note that at order $1/m_Q^2$ one has also a purely gluonic contribution  
due to closed heavy quark loops. It contains the gluonic field strength
tensor $G_{\mu\nu}$, which is defined by $i g_s G_{\mu\nu}=[D_\mu,D_\nu]$.

The first terms of these two expansions (\ref{FWfield})
define the static limit, 
which has been successfully applied to systems with a single heavy 
quark. In order to describe a system with more than one heavy (anti)quark
one has to write down the same expansion (\ref{FWfield}) for each heavy 
quark. However, there is no static limit for a system with two heavy quarks
if the two heavy quarks move with almost the same velocity as it is the 
case for a quarkonium; one runs into problems with diverging phases 
and ``complex anomalous dimensions'', which are considered in detail 
in \cite{KMO}. 

In order to cure this problem one has to choose the unperturbed 
system such that these phases are already generated by the leading order
dynamics, i.e.\ instead of the static limit one has to use the 
non-relativistic Lagrangian. In this case the ``power counting'' has to 
be modified in such a way that the leading term is the static plus the 
kinetic energy term, leading to the fact that the ``time derivative'' 
$ivD$ has to be counted as two powers of the ``spatial derivative''
$iD^\perp$. In physical terms, for a bound state this corresponds to 
the balance between potential and kinetic energy, since the covariant 
time derivative $ivD$ contains also the potential.  

For a system with a heavy quark and a heavy antiquark one then starts from 
\begin{equation} \label{lnull}
{\cal L}_0  = \bar{h}_v^{(+)} (ivD) h^{(+)}_v - \bar{h}_v^{(-)} (ivD) h^{(-)}_v
            + \bar{h}_v^{(+)} \frac{(iD_\perp)^2}{2m_Q} h^{(+)}_v
            + \bar{h}_v^{(-)} \frac{(iD_\perp)^2}{2m_Q} h^{(-)}_v
\end{equation}
where we have assumed for simplicity that the two quarks have the 
same mass; the case of unequal mass is obvious. 

Naive power counting would suggest to include also the Pauli term 
$\vec{\sigma} \cdot \vec{B}$ into the leading-order Lagrangian. However, 
in a coulombic system this term is in fact down by powers of $v_{rel}$, 
since the chromomagnetic field is generated by the relative motion of 
the two heavy quarks; this has been made manifest in \cite{LM}.

Most of the success of HQET is due to heavy quark flavor and spin symmetry.
However, once one uses (\ref{lnull}) the symmetries are somewhat different
for \HQQET or NRQCD. First of 
all, (\ref{lnull}) depends on the mass through the kinetic energy 
term; consequently 
the states will depend on $m_Q$ in a non-perturbative
way and heavy flavor symmetry is lost. In this way the mass dependent small 
scales (the inverse Bohr radius and the binding energy) 
are generated by taking 
matrix elements of operators involving derivatives with these states.   
On the other hand, since the chromomagnetic moment operator does not 
appear to leading order,  
(\ref{lnull}) does not depend on the spins of the two heavy quarks, and  
hence there is a spin symmetry which is larger than in HQET because we 
have two heavy quark spins; the resulting symmetry is an 
$SU(2) \otimes SU(2)$ corresponding to separate rotations of the 
two spins. 

For the case of heavy quarkonia all states fall into spin symmetry
quartets which should be degenerate in the non-relativistic limit. 
In spectroscopic notation  ${}^{2S+1}\ell_J$
these quartets consist of the states
\begin{equation}
[n{}^1 \ell_\ell \quad n{}^3 \ell_{\ell-1} \quad n{}^3 \ell_\ell \quad
 n{}^3 \ell_{\ell+1} ] .
\label{fourstates}
\end{equation}
For the ground states the spin symmetry quartet consists of the $\eta_Q$ 
(the $0^-$ state) and the three polarization directions of the 
$\Upsilon_Q$ (the $1^-$ state). 

The heavy quarkonia spin symmetry restricts the non-perturbative 
input to a calculation of processes involving heavy quarkonia. 
Of particular interest are decays in which the heavy quarks inside
the heavy quarkonium annihilate. The annihilation is a short 
distance process that can be calculated perturbatively in terms of 
quarks and gluons, while the long distance contribution is encoded in 
certain matrix elements of quark operators. Logarithmic dependences
on the heavy quark mass may be calculated by employing the usual 
renormalization group machinery.   

\section{Annihilation Decays of Heavy Quarkonia}
The starting point to calculate processes like 
$\eta_Q \to $ {\it light hadrons}, 
$\eta_Q \to \gamma \, + $ {\it light hadrons}, 
or the corresponding decays of the $\Upsilon_Q$ states
is the transition operator $T$ for two heavy quarks 
which annihilate into light degrees of freedom. This will in general 
be bilinear in the heavy quark fields, such that   
\begin{equation} \label{Top}
T (X,\xi) = (-i) \bar{Q}(X+\xi) K (X,\xi)  Q(X-\xi) \,,
\end{equation}
where $K (X,\xi)$ involves only light degrees of freedom and $X$ and $\xi$
correspond to the cms and relative coordinate respectively. If we identify 
the field $Q$ with the quark and $\bar{Q}$ with the antiquark, so 
we shall make the large scale $m_Q$ explicit by redefining the fields
as 
\begin{equation}
Q (x) = \exp(-im_Q (vx) ) Q_v^{(+)} (x) \,, \quad
\bar{Q} (x) =  \exp(-im_Q (vx) ) \bar{Q}_{v}^{(-)} (x) \,.
\end{equation}
This corresponds to the usual splitting of the heavy quark momentum 
into a large part $m_Q v$ and a residual piece $k$. Inserting this into 
(\ref{Top}) one finds
\begin{equation}
 T = (-i)  \exp[-i2 m_Q v X] 
     \bar{Q}_v^{(-)} (X+\xi) K (X,\xi)  Q_v^{(+)} (X-\xi) \,.
\end{equation}
The inclusive rate for the decay of a quarkonium 
$\Psi \to$ {\it light degrees of freedom} is then given by
\begin{equation}
\Gamma = \langle \Psi | 
\left[ \int d^4 X d^4 \xi d^4 \xi ' T (X,\xi) T^\dagger (0,\xi ') 
         + {\rm h.c.} \right] | \Psi \rangle \,.
\end{equation}
The next step is to perform 
an operator product expansion for the non-local product of the 
quark field operators. This expansion will yield four-quark operators 
of increasing dimension starting with dim-6 operators. The increasing 
dimension of these operators will be compensated by inverse powers of 
the heavy quark mass, so generically the rate takes the form
\begin {equation} \label{mexp}
\Gamma = m_Q \sum_{n,i} \left(\frac{1}{m_Q} \right)^{n-2} 
         C ({\cal O}_i^{(n)},\mu) \langle \Psi | 
         {\cal O}_i^{(n)}| \Psi \rangle|_\mu \,,
\end{equation}   
where $n=6,7,\ldots$ is the dimension of the operator and $i$ labels 
different operators with the same dimension. The coefficients  
$C ({\cal O}_i^{(n)},\mu)$ are related to the 
short distance annihilation process 
and hence may be calculated in perturbation theory in terms of quarks
and gluons. Once QCD radiative corrections are included, the 
$C ({\cal O}_i^{(n)},\mu)$ acquire a dependence on 
the renormalization scale 
$\mu$ which is governed by the renormalization group of the effective 
theory. The rate $\Gamma$ is independent of $\mu$ and hence the
$\mu$ dependence of $C ({\cal O}_i^{(n)},\mu)$ has to be compensated by 
a corresponding dependence of the matrix elements.     

The non-perturbative contributions are encoded in the 
matrix elements of the local four-quark operators, and the mass dependence 
of these operators is also expanded in powers of $1/m_Q$ and thus  the 
remaining $m_Q$ dependence of the matrix elements is only due to the 
states. In terms of the $m_Q$ independent static fields 
\begin{equation}
Q_v^{(+)} (x) = h_v^{(+)} (x) + {\cal O} (1/m_Q) \,, \quad
Q_v^{(-)} (x) = h_v^{(-)} (x) + {\cal O} (1/m_Q)
\end{equation}
one has in total four dim-6 operators
\[
A_1 ^{(C)} = [ \bar{h}_v^{(+)} \gamma_5 C h_v^{(-)} ] \, 
             [ \bar{h}_v^{(-)} \gamma_5 C h_v^{(+)} ] \mbox{ and }
A_2 ^{(C)} = [ \bar{h}_v^{(+)} \gamma^\perp_\mu C h_v^{(-)} ] \,
             [ \bar{h}_v^{(-)} \gamma_\perp^\mu C h_v^{(+)} ]
\]
with the color matrix $C$, where one has the two possibilities
$C \otimes C = 1 \otimes 1$ (color singlet) or
$C \otimes C = T^a \otimes T^a$ (color octet).
These operators do not mix under renormalization, all anomalous dimensions
vanish. 

There are no dim-7 operators, since these are either proportional to 
$(ivD)$ and can be rewritten in terms of dim-8 operators by the equations 
of motion (see (\ref{lnull})) or their forward matrix elements are forbidden
by symmetries. At dim-8 one finds 30 local operators
\begin{eqnarray*}
B_1^{(C)} &=& [ iD_\perp^\mu (\bar{h}_v^{(+)} \gamma_5 C h_v^{(-)}) ]
           \, [ iD^\perp_\mu (\bar{h}_v^{(-)} \gamma_5 C h_v^{(+)}) ]
\,, \\
B_2^{(C)} &=& [ iD_\perp^\mu (\bar{h}_v^{(+)} \gamma^\perp_\mu C h_v^{(-)}) ]
           \, [ iD_\perp^\nu (\bar{h}_v^{(-)} \gamma^\perp_\nu C h_v^{(+)}) ]
\,, \\
B_3^{(C)} &=& [ iD_\perp^\mu (\bar{h}_v^{(+)} \gamma_\perp^\nu C h_v^{(-)}) ]
           \, [ iD^\perp_\mu (\bar{h}_v^{(-)} \gamma^\perp_\nu C h_v^{(+)}) ]
\,, \\
\\
C_1^{(C)} &=& [ iD_\perp^\mu (\bar{h}_v^{(+)} \gamma_5 C h_v^{(-)}) ]
           \, [ \bar{h}_v^{(-)} \gamma_5
                 (\stackrel{\longleftrightarrow}{iD^\perp_\mu}) C h_v^{(+)} ]
           + {\rm h.c.} \,, \\
C_2^{(C)} &=& [ iD_\perp^\mu (\bar{h}_v^{(+)} \gamma^\perp_\mu C h_v^{(-)} ]
           \, [ \bar{h}_v^{(-)} 
           (\stackrel{\longleftrightarrow}{i\fmslash{D}^\perp}) C h_v^{(+)} ]
           + {\rm h.c.} \,, \\
C_3^{(C)} &=& [ iD_\perp^\mu (\bar{h}_v^{(+)} \gamma_\perp^\nu C h_v^{(-)}) ]
           \, [ \bar{h}_v^{(-)} \gamma^\perp_\nu
                 (\stackrel{\longleftrightarrow}{iD^\perp_\mu}) C h_v^{(+)} ]
           + {\rm h.c.} \,, \\
C_4^{(C)} &=& [ iD_\perp^\mu (\bar{h}_v^{(+)} \gamma_\perp^\nu C h_v^{(-)}) ]
           \, [ \bar{h}_v^{(-)} \gamma^\perp_\mu 
                 (\stackrel{\longleftrightarrow}{iD^\perp_\nu}) C h_v^{(+)} ]
           + {\rm h.c.} \,, \\
\\
D_1^{(C)} &=& [ \bar{h}_v^{(+)} \gamma_5
                 (\stackrel{\longleftrightarrow}{iD_\perp^\mu}) C h_v^{(-)} ]
           \, [ \bar{h}_v^{(-)} \gamma_5
                 (\stackrel{\longleftrightarrow}{iD^\perp_\mu}) C h_v^{(+)} ]
\,, \\
D_2^{(C)} &=& [ \bar{h}_v^{(+)}
           (\stackrel{\longleftrightarrow}{i\fmslash{D}^\perp}) C h_v^{(-)} ]
           \, [ \bar{h}_v^{(-)}
           (\stackrel{\longleftrightarrow}{i\fmslash{D}^\perp}) C h_v^{(+)} ]
\,, \\
D_3^{(C)} &=& [ \bar{h}_v^{(+)} \gamma_\perp^\mu
                 (\stackrel{\longleftrightarrow}{iD_\perp^\nu}) C h_v^{(-)} ]
           \, [ \bar{h}_v^{(-)} \gamma^\perp_\mu
                 (\stackrel{\longleftrightarrow}{iD^\perp_\nu}) C h_v^{(+)} ]
\,, \\
D_4^{(C)} &=& [ \bar{h}_v^{(+)} \gamma_\perp^\mu
                 (\stackrel{\longleftrightarrow}{iD_\perp^\nu}) C h_v^{(-)} ]
           \, [ \bar{h}_v^{(-)} \gamma^\perp_\nu
                 (\stackrel{\longleftrightarrow}{iD^\perp_\mu}) C h_v^{(+)} ]
\,, \\
\\
E_1^{(C)} &=& [ \bar{h}_v^{(+)} \gamma_5 C h_v^{(-)}] \,
              [ \bar{h}_v^{(-)} \gamma_5
                   (\stackrel{\longleftrightarrow}{iD^\perp})^2 C h_v^{(+)} ]
           + {\rm h.c.} \,, \\
E_2^{(C)} &=& [ \bar{h}_v^{(+)} \gamma_\perp^\mu C h_v^{(-)}] \, 
              [ \bar{h}_v^{(-)}
		 (\stackrel{\longleftrightarrow}{i\fmslash{D}^\perp})
                 (\stackrel{\longleftrightarrow}{iD^\perp_\mu}) C h_v^{(+)} ]
           + {\rm h.c.} \,, \\
E_3^{(C)} &=& [ \bar{h}_v^{(+)} \gamma_\perp^\mu C h_v^{(-)}] \, 
              [ \bar{h}_v^{(-)} (\stackrel{\longleftrightarrow}{iD^\perp_\mu}) 
           (\stackrel{\longleftrightarrow}{i\fmslash{D}^\perp}) C h_v^{(+)} ]
           + {\rm h.c.} \,, \\
E_4^{(C)} &=& [ \bar{h}_v^{(+)} \gamma_\perp^\mu C h_v^{(-)}] \, 
              [ \bar{h}_v^{(-)} \gamma^\perp_\mu
                   (\stackrel{\longleftrightarrow}{iD^\perp})^2 C h_v^{(+)} ]
           + {\rm h.c.} \,,
\end{eqnarray*}
where we have defined 
\begin{equation}
\bar{f}(x) (\stackrel{\leftrightarrow}{D}) g(x)
= \bar{f}(x) (Dg(x)) - (\overline{Df(x)}) g(x)  
\end{equation}
corresponding to (the gauge invariant generalization of) the
relative momentum.
 
Note that we have split these operators into four blocks. The 
$B_i^{(C)}$ operators contain two derivatives acting on both 
the heavy quark and the heavy  antiquark 
operators; thus they correspond to a motion of the cms-system of the 
two heavy quarks relative to the cms-system of the quarkonium. The 
operators of the $C_i^{(C)}$ type are chosen such that one derivative 
acts again on both fields, while the other derivative corresponds to 
the relative motion of the heavy quark-antiquark pair. Finally, the 
$D_i^{(C)}$ and $E_i^{(C)}$ correspond to two powers of the 
relative momentum of the two heavy quarks. 

\section{Renormalization Group Evolution}
In (\ref{mexp}) a spurious dependence on the renormalization point $\mu$
appears, where $\mu$ separates long from short distances. The coefficients 
in front of the matrix elements may be calculated in perturbation theory 
and are determined at the scale $\mu = m_Q$ from a matching calculation. 
Once the matching is done one has to evolve down to a small scale which 
is typical for the matrix elements appearing in (\ref{mexp}). As it was 
discussed in the introduction there are several small scales in a quarkonium
and in this context several questions arise. Firstly, if the dimension 
is not the correct way of counting the relevance of a matrix element, how is 
the counting of powers done correctly? Secondly, how far should one 
evolve down using the renormalization group in order to get the ``best''
result?

For superheavy quarkonia, i.e.\ the ones one may obviously describe using 
NRQCD, these problems are easier to discuss. Since the binding is due to 
instantaneous coulomb gluons, one may write the Lagrangian in Coulomb 
gauge as 
\begin{equation}
{\cal L} = \bar{h}_v^{(+)} (ivD) h^{(+)}_v - \bar{h}_v^{(-)} (ivD) h^{(-)}_v 
         + \bar{h}_v^{(+)} \frac{(i\partial^\perp)^2}{2m_Q} h^{(+)}_v
         + \bar{h}_v^{(-)} \frac{(i\partial^\perp)^2}{2m_Q} h^{(-)}_v
\end{equation}
since the gluon fields associated with the spatial derivatives $iD^\perp$ 
are suppressed due to the smallness of the coupling 
$\alpha_s (m_Q v_{rel}^2)$. Furthermore, coulomb gluons do not appear as 
dynamical degrees of freedom, rendering the superheavy quarkonium as a 
real non-relativistic two particle state, with a potential determined
from the gluon Greens function 
$\left\langle 0 | A_0 (x) A_0 (y) | 0 \right\rangle$. 
In such a case one may perform a Fock state decomposition \cite{BBL}
of the quarkonium of the form
\begin{equation} \label{fock}
| \Psi \rangle = C_0 | \bar{Q} Q \rangle + 
                 C_1 | \bar{Q} Q g \rangle + 
                 C_2 | \bar{Q} Q gg \rangle + \cdots 
\end{equation}
in which subsequent coefficients are suppressed by powers of $v_{rel}$, 
$C_i \sim v_{rel}^i$, since each dynamical gluon is suppressed by a power of 
$v_{rel}$. In addition, one has the same selection rules as in 
Non-Relativistic QED (NRQED), yielding a consistent scheme for an expansion
in $v_{rel}$. 

In this way to count powers the matrix elements of the operators  
$B_i^{(C)}$ and $C_i^{(C)}$
are down by additional powers of $v_{rel}$ compared to their
dimension, while the operators of the $D_i^{(C)}$ and $E_i^{(C)}$ type
are of order $(m_Q v_{rel})^2$ and thus their matrix elements constitute 
the first subleading corrections in the power counting scheme along the lines 
of NRQED. The same argument applies in for the operators in which the two 
quarks are in a color octet state. Since the quarkonium has to form a 
color singlet, these contributions have to be of the form of a higher Fock
component and are thus suppressed by additional powers of $v_{rel}$.

\begin{figure}
  \begin{center}
  \leavevmode
    \epsfxsize=14cm
    \epsffile[0 110 610 295]{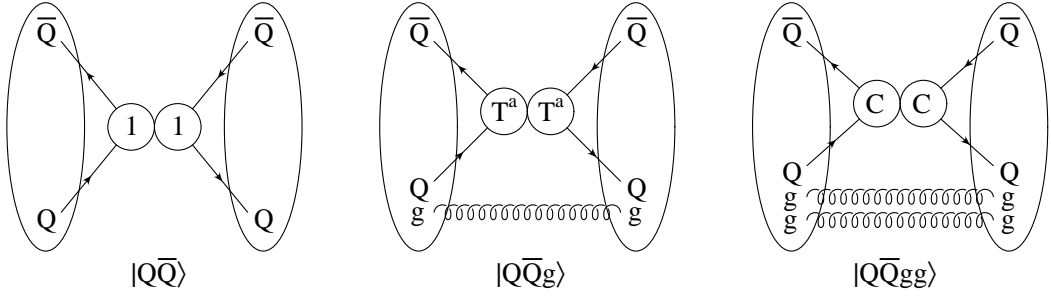}
  \end{center}
  \centerline{\parbox{14cm}{\caption{\label{HQQETillu}
              Illustration of the various Fock components. In the standard
              NRQCD picture the left figure represents the leading
              contribution.}}}
\end{figure}

However, realistic quarkonia seem not to be close to this limit, and it is 
well possible, that the $v_{rel}$ counting scheme fails for these systems. 
In such a case it seems to be useful to count the operators according to 
their dimension given by the powers of $iD^\perp$. This means 
that all the matrix elements of the operators $B_i^{(C)}, ... , E_i^{(C)}$
will be taken as the subleading terms. In particular this means that the
states of the effective theory are not simply non-relativistic two particle
states in which the two heavy quarks are bound by a potential generated by 
instantaneous, coulomb type gluons. Rather they have to contain dynamical
gluons as well such that all the Fock components in (\ref{fock}) are of
comparable size. The physical picture is illustrated in fig.\ref{HQQETillu}. 

The operators  $B_i^{(C)}, ... , E_i^{(C)}$ are all local operators
corresponding to the $1/m_Q$ expansion of the hard kernel.  
In addition to these local contributions one has 
also non-local terms originating
from single insertions of the Lagrangian of order $1/m_Q^2$ and from 
double insertions of the Lagrangian of order $1/m_Q$. These contributions 
correspond to the corrections to the states of the effective theory 

Under 
renormalization the local dim-8 operators do not mix; only the double 
insertion of the kinetic energy operator of order $1/m_Q$
($K_1 = K_1^{(+)} + K_1^{(-)}$) mixes into some of the above operators.
Denoting this contribution as
\begin{equation}
T^{(C)}_i = \frac{(-i)^2}{2} \int\!\! d^4x d^4y \, T[A_i^{(C)} K_1(x) K_1 (y)]
\end{equation}
one obtains in one-loop renormalization group improved perturbation theory
two sets of equations for the coefficients of the operators with the 
spin structure $\gamma_5 \otimes \gamma_5$ \cite{BBL, MS}
\begin{eqnarray}
C(D_1^{(1)},\mu) &=& C(D_1^{(1)},m_Q) + \frac{32}{9} 
                     \frac{1}{33-2n_f} \, C(T_1^{(8)},m_Q) \ln \eta \,,
\nonumber \\
C(E_1^{(1)},\mu) &=& C(E_1^{(1)},m_Q) - 
                     \frac{8}{33-2n_f} \, C(T_1^{(1)},m_Q) \ln \eta \,,
\nonumber \\
C(B_1^{(8)},\mu) &=& C(B_1^{(8)},m_Q) - 
                     \frac{24}{33-2n_f} \, C(T_1^{(8)},m_Q) \ln \eta \,,
\nonumber \\
C(D_1^{(8)},\mu) &=& C(D_1^{(8)},m_Q) + 
                     \frac{16}{33-2n_f} \, C(T_1^{(1)},m_Q) \ln \eta \,,\\
                  && \hspace{2.2cm} \mbox{}+
           \frac{20}{3}\frac{1}{33-2n_f} \, C(T_1^{(8)},m_Q) \ln \eta \,,
\nonumber \\
C(E_1^{(8)},\mu) &=& C(E_1^{(8)},m_Q) - \frac{14}{3}
                     \frac{1}{33-2n_f} \, C(T_1^{(1)},m_Q) \ln \eta \,,
\nonumber
\end{eqnarray}
where $n_f$ is the number of active flavors and
$\eta = (\alpha_s (\mu) / \alpha_s (m_Q) )$. Furthermore, the 
coefficients $C(T_i^{(C)},m_Q)$ are the same as the ones for the dim-6 
operators $C(A_i^{(C)},m_Q)$ since the kinetic energy operator is not 
renormalized.

The second set of equations is for the operators with spin structure
$\gamma_\mu \otimes \gamma^\mu$ and due to heavy quarkonia spin symmetry 
one obtains the same equations; all other renormalization
group equations are trivial. 

The calculation of the coefficients is based on perturbation theory, and 
hence $\alpha_s$ has to be sufficiently small. Although the final result 
does not depend on $\mu$, there is still the practical question of how far 
one should run using the renormalization group. Clearly $\mu$ should be 
chosen to be one of the three, the inverse Bohr radius, the binding energy or 
$\Lambda_{QCD}$. In the worst case $\Lambda_{QCD}$ is of the order of the 
inverse Bohr radius, and then the renormalization group evolution has to 
stop there. In the superheavy case, $\Lambda_{QCD}$ is small compared to
the binding energy and one may run down to the scale set by the binding 
energy, below this scale one would need to switch to an effective theory
involving hadronic instead of QCD degrees of freedom.

\section{Hadronic Matrix Elements and Spin Symmetry}
A calculation of an annihilation decay then involves to calculate the 
$C({\cal O}_i^{(n)},\mu)$ at the scale $\mu = m_Q$ by 
matching the effective theory to full QCD. Once this is done, one may 
run down to some small scale $\mu$. In the superheavy case $\mu$ is 
of the order of the binding energy
of the heavy quarkonium, thereby resumming the well known logarithms of the 
form $\ln(m_Q / \mu)$ that appear in the calculations of decay rates of 
heavy $p$-wave quarkonia. As an example we shall study 
the decay $\eta_Q \to \gamma$ + {\it light hadrons} in the next section 
and consider the first nontrivial corrections to this mode. 

The matrix elements of the operators $B_i^{(C)}, ... , E_i^{(C)}$
as well as the non-local terms are non-perturbative quantities, 
which are constrained by heavy quarkonia spin symmetry. 
In order to exploit this symmetry, one may use the 
usual representation matrices for the spin singlet and spin triplet 
quarkonia
\begin{equation}
H_1 (v) = \sqrt{M} P_+ \gamma_5 \mbox{ for } S = 0 \,, \quad
H_3 (v) = \sqrt{M} P_+ \epsilon \mbox{ for } S = 1 
\end{equation}
where $M \approx 2m_Q$ is the mass of the heavy quarkonium and 
$P_+ = (1+\fmslash{v})/2$.   
Using this one finds for the matrix elements 
of the dim-6 operators
\[
\langle \Psi | \bar{h}^{(+)} \Gamma C h^{(-)}\, 
               \bar{h}^{(-)} \Gamma ' C h^{(+)}  | \Psi \rangle
= a^{(C)} (n,\ell) G \, \mbox{ with }
 G = {\rm Tr} (\overline{H}_{2s+1} \Gamma ) {\rm Tr} (\Gamma ' H_{2s+1} ) 
\]
Thus for each $n$ and $\ell$ and for each color combination one finds 
a single parameter for both the spin singlet and spin triplet quarkonium. 

Correspondingly for the dim-8 operators we get 
\begin{eqnarray*}
\langle\Psi| [ iD^\perp_\mu (\bar{h}_v^{(+)} \Gamma C h_v^{(-)}) ] \, 
           [ iD^\perp_\nu (\bar{h}_v^{(-)} \Gamma' C h_v^{(+)}) ] |\Psi \rangle
&=& b^{(C)}(n,\ell) (g_{\mu \nu} - v_\mu v_\nu) G
\\
\langle\Psi| [ \bar{h}_v^{(+)} \Gamma
               (\stackrel{\longleftrightarrow}{iD^\perp_\mu}) C h_v^{(-)} ] \,
             [ iD^\perp_\nu (\bar{h}_v^{(-)} \Gamma' C h_v^{(+)}) ]
                                                      + {\rm h.c.} |\Psi\rangle
&=& c^{(C)}(n,\ell) (g_{\mu \nu} - v_\mu v_\nu) G
\\
\langle\Psi| [ \bar{h}_v^{(+)} \Gamma
               (\stackrel{\longleftrightarrow}{iD^\perp_\mu}) C h_v^{(-)} ] \,
             [ \bar{h}^{(-)} \Gamma'
               (\stackrel{\longleftrightarrow}{iD^\perp_\nu}) C h_v^{(+)} ] \,
&=& d^{(C)}(n,\ell) (g_{\mu \nu} - v_\mu v_\nu) G
\\
\langle\Psi| [ \bar{h}^{(+)} \Gamma
               (\stackrel{\longleftrightarrow}{iD^\perp_\mu})
               (\stackrel{\longleftrightarrow}{iD^\perp_\nu}) C h_v^{(-)} ] \,
             [ \bar{h}^{(-)} \Gamma' C h^{(+)}] + {\rm h.c.} |\Psi\rangle
&=& e^{(C)}(n,\ell) (g_{\mu \nu} - v_\mu v_\nu) G 
\end{eqnarray*}
For fixed values of $n$ and $\ell$ one finds that eight parameters are needed 
to describe the matrix elements of the dim-8 operators. 

These matrix elements are non-perturbative, but from vacuum insertion one 
is lead to assume 
\begin{eqnarray} 
a^{(1)}(n,0) = \frac{3}{8\pi} |R_{n0}(0)|^2
             &\gg& a^{(1)}(n,\ell) \mbox{ for } \ell \ne 0 \, , \nonumber \\
d^{(1)}(n,1) = \frac{3}{2\pi} |R'_{n1}(0)|^2
             &\gg& d^{(1)}(n,\ell) \mbox{ for } \ell \ne 1 \, , \\
e^{(1)}(n,0) = \frac{6}{\pi} \, {\rm Re} [R''_{n0}(0)R^*_{n0}(0)] 
             &\gg& e^{(1)}(n,\ell) \mbox{ for } \ell \ne 0 \, , \nonumber
\end{eqnarray}
\begin{eqnarray}
a^{(1)}(n,0) &\gg& a^{(8)}(n,\ell) \mbox{ for all } n, \ell \, , \nonumber \\
d^{(1)}(n,1) &\gg& d^{(8)}(n,\ell) \mbox{ for all } n, \ell \, , \\
e^{(1)}(n,0) &\gg& e^{(8)}(n,\ell) \mbox{ for all } n, \ell \, , \nonumber
\end{eqnarray}
where $R_{nl} (r)$ is the radial wave function of the quarkonium. The same 
reasoning 
yields the expectation that $b^{(C)} (n,\ell)$ and $c^{(C)} (n,\ell)$ are 
small compared to the coefficients that are non-vanishing in vacuum insertion.

In fact, the expectations from vacuum insertion are strengthened by the 
superheavy case in which vacuum insertion is true to leading order in the 
$v_{rel}$ expansion. Furthermore, the color octet as well as the operators
of the $B_i^{(C)}$ and $C_i^{(C)}$ type are also down by powers of $v_{rel}$.

\section{The Photon Spectrum in $\eta_Q \to \gamma$ + {\it light hadrons}}
\begin{figure}
  \begin{center}
  \leavevmode
    \epsfysize=5cm
    \epsffile{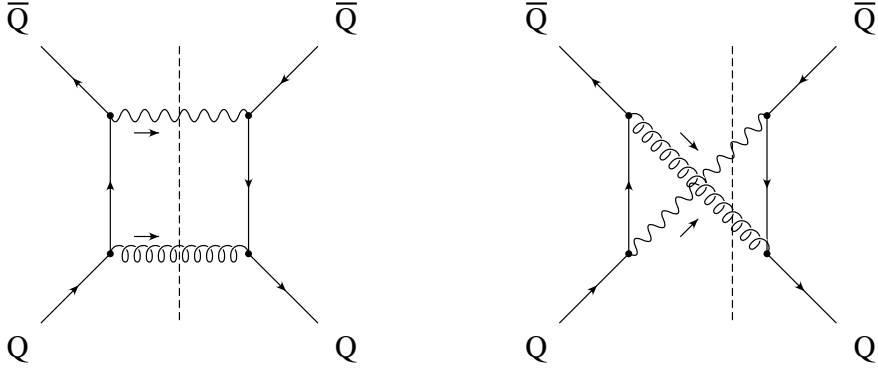}
  \end{center}
  \centerline{\parbox{16cm}{\caption{\label{fig1}
The hard subprocesses for $\eta_Q \to \gamma$ + {\it light hadrons}.
}}}
\end{figure}

The short distance part of the process under consideration is given 
by the amplitude obtained from the Feynman diagrams depicted in 
fig.\ref{fig1}. The amplitudes are evaluated in QCD and then matched
to \HQQET by performing an expansion in $1/m_Q$. At the matching scale 
the spectrum, including the first non-trivial corrections, has the form
\begin{equation} \label{spectrum}
\frac{d\Gamma}{dx} =  {\cal G}
\left[\delta(1-x){\cal A} + \frac{1}{4m_Q^2} 
\left\{\delta(1-x){\cal B}_0 + \delta^\prime (1-x){\cal B}_1
 + \delta^{\prime \prime} (1-x) {\cal B}_2 \right\}\right]
\end{equation}
with 
\begin{equation} 
{\cal G} = \frac{8\pi \alpha_s(m_Q) \alpha_{em} Q_Q^2} {m_Q^2} \,,
\end{equation}
where $x := E_\gamma/m_Q$ is the scaled photon energy, $\alpha_{em} = 1/137$
is the electromagnetic coupling and $Q_Q$ is the charge of the heavy quark.  
At the matching scale $\mu = m_Q$ the coefficients of the $\delta$-functions
in (\ref{spectrum}) are given in terms of the forward matrix elements discussed
in the last section and non-local terms originating from the correction in 
the Lagrangian. One finds from the matching calculation
\begin{eqnarray}
{\cal A} &=& \langle \eta_Q | A_1^{(8)} | \eta_Q \rangle |_{m_Q} \\
{\cal B}_0 &=&  
   \frac{4}{3} \langle \eta_Q | E_1^{(8)} | \eta_Q \rangle |_{m_Q}
 - \frac{7}{3} \langle \eta_Q | B_1^{(8)} | \eta_Q \rangle |_{m_Q} \nonumber \\
 && + (-i) \int d^4x \, \langle \eta_Q | 
     T\{ ( K_2 (x)+{\cal L}_{glue}(x) ) A_1^{(8)}\} | \eta_Q \rangle |_{m_Q} \\
 && + \frac{(-i)^2}{2} \int d^4x \, d^4 y \,  
   \langle \eta_Q | T\{ M_1 (x) M_1 (y) A_1^{(8)}\} | \eta_Q \rangle |_{m_Q} 
\nonumber \\
{\cal B}_1 &=& 
     \frac{1}{2} \langle \eta_Q | B_1^{(8)} | \eta_Q \rangle |_{m_Q}
   - \frac{1}{4} \langle \eta_Q | E_1^{(8)} | \eta_Q \rangle |_{m_Q} \\
{\cal B}_2 &=& \frac{1}{6} \langle \eta_Q | B_1^{(8)} | \eta_Q \rangle |_{m_Q}
\end{eqnarray}
The renormalization group may now be used to run down to some small scale 
$\mu \sim \Lambda_{QCD}$, thereby extracting logarithms of the form
$\log (m_Q / \mu)$ from the matrix elements. Using the one loop result 
as given in the last section, one has at the small scale
\begin{eqnarray}
{\cal A} &=& \langle \eta_Q | A_1^{(8)} | \eta_Q \rangle |_\mu \\
{\cal B}_0 &=& 
   \frac{1}{3} \left[4 - \frac{56}{33 - 2n_f} \ln \eta \right]
\langle \eta_Q | E_1^{(8)} | \eta_Q \rangle |_\mu
 - \frac{1}{3} \left[ 7 - \frac{288}{33 - 2n_f} \ln \eta \right]
\langle \eta_Q | B_1^{(8)} | \eta_Q \rangle |_\mu \nonumber \\
&& + \frac{128}{3} \frac{1}{33 - 2n_f} \ln \eta 
\langle \eta_Q | D_1^{(1)} | \eta_Q \rangle |_\mu 
+ \frac{1}{33 - 2n_f} \ln \eta 
\langle \eta_Q | D_1^{(8)} | \eta_Q \rangle |_\mu \nonumber \\
&& + (-i) \int d^4x \, \langle \eta_Q | 
   T\{ ( K_2 (x)+{\cal L}_{glue}(x) ) A_1^{(8)}\} | \eta_Q \rangle |_\mu \\
&& + \frac{(-i)^2}{2} \int d^4x \, d^4 y \,  
\langle \eta_Q | T\{ (M_1 (x) M_1 (y) A_1^{(8)}\} | \eta_Q \rangle |_\mu
\nonumber \\
{\cal B}_1 &=& 
    \frac{1}{2} \langle \eta_Q | B_1^{(8)} | \eta_Q \rangle |_\mu
  - \frac{1}{4} \langle \eta_Q | E_1^{(8)} | \eta_Q \rangle |_\mu \\
{\cal B}_2 &=& \frac{1}{6} \langle \eta_Q | B_1^{(8)} | \eta_Q \rangle |_\mu
\end{eqnarray}
Note that the renormalization group flow has induced two operators that 
have not been present at the matching scale. These type of logarithms 
has been observed already in the calculation of $p$-wave quarkonia 
some time ago \cite{Gatto}; 
these have been fixed order calculation and the logarithms 
appear here as infrared singularities. In the effective theory approach they 
are generated by the renormalization group flow which in addition 
even resums these terms, since 
\begin{eqnarray}
- \ln \eta &=& \ln \left(1 + \alpha_s(m_Q)
                        \frac{\beta_0}{2\pi} \ln \frac{\mu}{m_Q} \right) 
\nonumber \\
        &=& \alpha_s(m_Q) \frac{\beta_0}{2\pi} \ln \frac{\mu}{m_Q}
        + \frac{1}{2} \left( \alpha_s(m_Q)
                             \frac{\beta_0}{2\pi} \ln \frac{\mu}{m_Q} \right)^2
        + \ldots \,,
\end{eqnarray}
where $\beta_0=11-2n_f$.

The expression obtained for the photon spectrum contains $\delta$-function 
and its derivatives which is obviously unphysical. The origin of this 
singular behavior is of kinematical nature: At the partonic 
level, the initial state quarks 
move with the same velocity and hence act like a single particle, which 
then decays into two massless objects. Thus this is a two particle decay 
and so the energies of the final state particles are fixed. 

In order to compare with the observed hadron spectrum one has to apply 
some ``smearing'' in the sense of \cite{WPQ}; in particular, if one 
calculates moments of the spectrum (\ref{spectrum}) one obtains a 
sensible answer which may be compared to the observed spectrum\footnote{
  In fact,
  the situation is completely the same as in the endpoint of inclusive 
  $B \to X_u \ell \nu$ or $B \to X_s \gamma$ decays \cite{neubert},
  where the $1/m_Q$ expansion also only yields the moments of the observed 
  spectrum.}.
In other words the $1/m_Q$ expansion yields an expansion of the spectrum in 
terms of singular functions 
\begin{equation}
\frac{d\Gamma}{dx} = \sum_{n=0}^\infty \frac{1}{n!} M_n \delta^{(n)} (1-x) \,,
\end{equation}
where the coefficients of the expansion are the moments 
\begin{equation}
M_n = \int_0^1 dx  \frac{d\Gamma}{dx} (1-x)^n \,.
\end{equation}
The zeroth moment is simply the total rate 
\begin{equation}
\Gamma = {\cal G} [{\cal A} + \frac{1}{4m_Q^2} {\cal B}_0] \,,
\end{equation}
while the first and the second moments are entirely of order $1/m_Q^2$
\begin{equation}
M_1 = {\cal G} {\cal B}_1 \,, \qquad M_2 = {\cal G} {\cal B}_2 \,.
\end{equation}
As expected, the first and the second moment are entirely proportional to 
color octet contributions, while the renormalization group flow induces
the color singlet operator $D_1^{(1)}$. In factorization as well as in 
the NRQCD case all these operators are down by at least two powers of 
$v_{rel}$ and hence this process would be very much suppressed compared to 
allowed processes such as $J/\Psi \to \gamma$ + {\it light hadrons}. Hence
this process is a nice testing ground for factorization and the power
counting scheme of NRQCD. 

\section{Conclusions}
Unlike systems with a single heavy quark quarkonia-like systems are much more 
difficult to describe. The simplicity of HQET is due to the fact that only 
a single small scale appears, which is independent of the heavy mass, and 
thus a static limit may be performed. In a quarkonium several small scales
appear which depend in a non-perturbative way on the heavy mass. The 
superheavy case resembles very much a QED like system which is bound
by coulombic forces. In such a system one has the inverse Bohr radius 
$m_Q v_{rel}$ and the binding energy $m_Q v_{rel}^2 $ as small scales, and 
if $\Lambda_{QCD} \ll m_Q v_{rel}^2 $ the binding is in the sense perturbative
that it is due to one gluon exchange which takes in coulomb gauge the form 
of an instantaneous potential. 

In this superheavy case one may transfer practically all the knowledge of 
the description of positronium in NRQED to the QCD case. However, realistic 
quarkonia do not seem to be too close to this limit, since the charm 
and the bottom mass are too small to fulfill $\Lambda_{QCD} \ll m_Q v_{rel}^2$.
Thus the power counting scheme of NRQED cannot be naively transferred to QCD,
and the way how to organize the effective theory calculation becomes obscure. 

The safe way in this case is to rely only on the dimension of the 
operators involved, thereby taking into account a proliferation of unknown 
parameters, which all would need to be determined from experiment. It is 
then of some interest to define observables which are sensitive to specific 
matrix elements and hence may shed some light on what happens in realistic     
quarkonia. 

In the present paper we have studied the decay $\eta_Q \to \gamma$ + {\it 
light hadrons}, which would be strongly suppressed compared to $J/\Psi \to
\gamma$ + {\it light hadrons}, if both $J/\Psi$ and $\eta_Q$ were superheavy.
However, this process is not suppressed due to higher powers of $1/m_Q$
appearing in the heavy mass expansion, rather it is suppressed due to the
quarkonia states, with which the matrix elements are taken. 

It turns out that the moments of the measured photon spectrum in 
$\eta_Q \to \gamma$ + {\it light hadrons} is sensitive to matrix elements
which would be strongly suppressed for the superheavy case. Unfortunately,
there are not yet data on this process, such that a check, whether it is
really suppressed as predicted by the counting of powers of $v_{rel}$ has
to wait for future experiments. 

\section*{Acknowledgments}
We would like to thank Martin Beneke and Gerhard Schuler for
discussions and Andreas Schenk for checking details of the calculations. 
This work was supported by Deutsche Forschungsgemeinschaft.


\begin{thebibliography}{99}
\bibitem{SV87} {M. Voloshin and M. Shifman,
               Sov. J. Nucl. Phys. {\bf 45} (1987) 292  and
               {\bf 47} (1988) 511;\hfill\\
               N. Isgur and M. Wise,
               Phys. Lett. {\bf B232} (1989) 113 and
               {\bf B237} (1990) 527;\hfill\\
               E. Eichten and B. Hill,
               Phys.\ Lett.\ {\bf B234} (1990) 511;\hfill\\
               B. Grinstein,
               Nucl. Phys. {\bf B339} (1990) 253;\hfill\\
               H. Georgi,
               Phys. Lett. {\bf B240} (1990) 447; \hfill\\
               A. Falk, H. Georgi, B. Grinstein and M. Wise,
               Nucl. Phys. {\bf B343} (1990) 1.}
\bibitem{reviews}{H. Georgi: contribution to the
               {\it Proceedings of TASI--91},
               by R.K.\ Ellis et al. (eds.)
               (World Scientific, Singapore, 1991);\hfill\\
               B. Grinstein: contribution to {\it High Energy
               Phenomenology},
               R. Huerta and M.A.\ Peres (eds.)
              (World Scientific, Singapore, 1991);\hfill\\
               N. Isgur and M. Wise: contribution to
               {\it Heavy Flavors},
               A. Buras and M. Lindner (eds.)
              (World Scientific, Singapore, 1992);\hfill\\
               M. Neubert, SLAC--PUB 6263 (1993)
               (to appear in Phys. Rep.);
               \hfill\\
               T. Mannel,  contribution to {\it QCD--20 years later},
               P. Zerwas and H. Kastrup (eds.)
              (World Scientific, Singapore, 1993). }
\bibitem{KMO} {B. Grinstein, W. Kilian, T. Mannel and M. Wise,
                Nucl.\ Phys.\ {\bf B363} (1991)  19;\hfill\\
               W. Kilian, P. Manakos and T. Mannel,
               Phys.\ Rev.\ {\bf D48} (1993) 1321;\hfill\\
               W. Kilian, T. Mannel and T. Ohl,
               Phys.\ Lett.\ {\bf B304} (1993) 311.}
\bibitem{BBL} {G.T. Bodwin, E. Braaten and G.P. Lepage,
               Phys.\ Rev.\ {\bf D51} (1995) 1125.}
\bibitem{MS} {T. Mannel and G. Schuler, Z. Phys.\ {\bf C67} (1995) 159;\hfill\\
              T. Mannel and G. Schuler, Phys.\ Lett.\ {\bf B349} (1995) 181.}
\bibitem{MRR} {T. Mannel, W. Roberts and Z. Ryzak,
               Nucl.\ Phys.\ {\bf B368} (1992) 204.}
\bibitem{FW}  {J. K\"orner and G. Thompson,
               Phys.\ Lett.\ {\bf B264} (1991) 185;\hfill\\
               S. Balk, F. K\"orner and D. Pirjol,
               Nucl.\ Phys.\ {\bf B428} (1994) 499.}
\bibitem{LM}  {M. Luke and A. V. Manohar, UTPT 96--14, UCSD/PTH 96--24,
               hep--ph/9610534}
\bibitem{Gatto}{R. Barbieri, R. Gatto and E. Remiddi,
               Phys.\ Lett.\ {\bf B61} (1976) 465.}
\bibitem{WPQ} {E.C. Poggio, H.R. Quinn and S. Weinberg,
               Phys.\ Rev.\ {\bf D13} (1976) 1958.}
\bibitem{neubert}{M. Neubert, Phys.\ Rev.\ {\bf D49} (1994) 3392;\hfill\\
                  M. Neubert, Phys.\ Rev.\ {\bf D49} (1994) 4623;\hfill\\
                  I.I. Bigi, M.A. Shifman, N.G. Uraltsev and A.I. Vainshtein,
                  Phys.\ Rev.\ {\bf D52} (1995) 196.}
\end{thebibliography}
\end{document}